\documentclass{PoS}
\usepackage{url}
\usepackage{cite}
\usepackage{subcaption}

\title{Multi-stage jet evolution through QGP using the JETSCAPE framework: inclusive jets, correlations and leading hadrons}

\ShortTitle{Multi-stage jet evolution through QGP ...}

\author{\speaker{Chanwook Park} for the JETSCAPE collaboration\\
	Department of physics, McGill University\\
	3600 University Street, Montr\'{e}al, QC, H3A 2T8, Canada\\
        E-mail: \email{chanwook@physics.mcgill.ca}}

\abstract{
The JETSCAPE Collaboration has recently announced the first release of the JETSCAPE package that provides a modular, flexible, and extensible Monte Carlo event generator.
This innovative framework makes it possible to perform a comprehensive study of multi-stage high-energy jet evolution in the Quark-Gluon Plasma.
In this work, we illustrate the performance of the event generator for different algorithmic approaches to jet energy loss, and reproduce the measurements of several jet and hadron observables as well as correlations between the hard and soft sector.
We also carry out direct comparisons between different approaches to energy loss to study their sensitivity to those observables. 
}

\FullConference{International Conference on Hard and Electromagnetic Probes of High-Energy Nuclear Collisions\\
		30 September - 5 October 2018\\
		Aix-Les-Bains, Savoie, France}

\begin{document}

\section{Introduction}
The JETSCAPE (Jet Energy-loss Tomography with a Statistically and Computationally Advanced Program Envelope) collaboration has recently released a modular, flexible, and extensible Monte Carlo event generator~\cite{Kauder:2018cdt} to the heavy-ion community.
High-energy jet evolution through a hot and dense QCD plasma is a multi-scale problem, with several distinct stages.
Different theoretical descriptions of the evolution should be utilized to correctly describe the parton shower at different stages, depending on the energy and virtuality of the jet.

The JETSCAPE framework was benchmarked by our previous work~\cite{Cao:2017zih}, where a comparison was made between different energy loss models for jet propagating through a brick. 
This framework provides a unified approach, where stages can be calculated using the appropriate models. 
In this proceedings, we present a selection of first results obtained from this comprehensive modeling approach.
The focus of this work is mainly on jet and leading hadron dynamics (inclusive suppression, and correlation between hard and soft sectors) at the LHC. 
Comparisons between different energy loss approaches are carried out to explore the effect of each approach in the full history of parton evolution in heavy-ion collisions.

\section{Unified approach in JETSCAPE}

In the current version of the JETSCAPE framework, four different energy loss modules are available; MATTER~\cite{Majumder:2013re} for modeling high virtuality evolution, $Q^2 \gg \sqrt{\hat{q}E}$; LBT~\cite{Cao:2017hhk}, MARTINI~\cite{Schenke:2009gb}, and AdS/CFT~\cite{Casalderrey-Solana:2014bpa} for modeling a low virtuality parton shower.
Initial hard partons generated by PYTHIA~\cite{Sjostrand:2007gs} are fed into MATTER and propagated with a virtuality-ordered shower in vacuum or medium.
If the virtuality of a given parton drops below a specified separation scale $Q_0$, the low-virtuality energy loss modules take the parton over for time-ordered evolution.
Switching between different energy loss modules is done independently for each parton. 
The separation scale $Q_0$ is set to $2$ GeV.
Hadronization is performed by the Lund-model-based fragmentation model, where no color information for each parton is tracked.
The event-averaged hydrodynamic background is provided by 2+1D VISHNU~\cite{Shen:2014vra} with T$\raisebox{-0.4ex}{R}$ENTo~\cite{Moreland:2014oya} initial conditions.

\section{Results}

\begin{figure*}[t]
	\centering
	\includegraphics[width=0.9\textwidth]{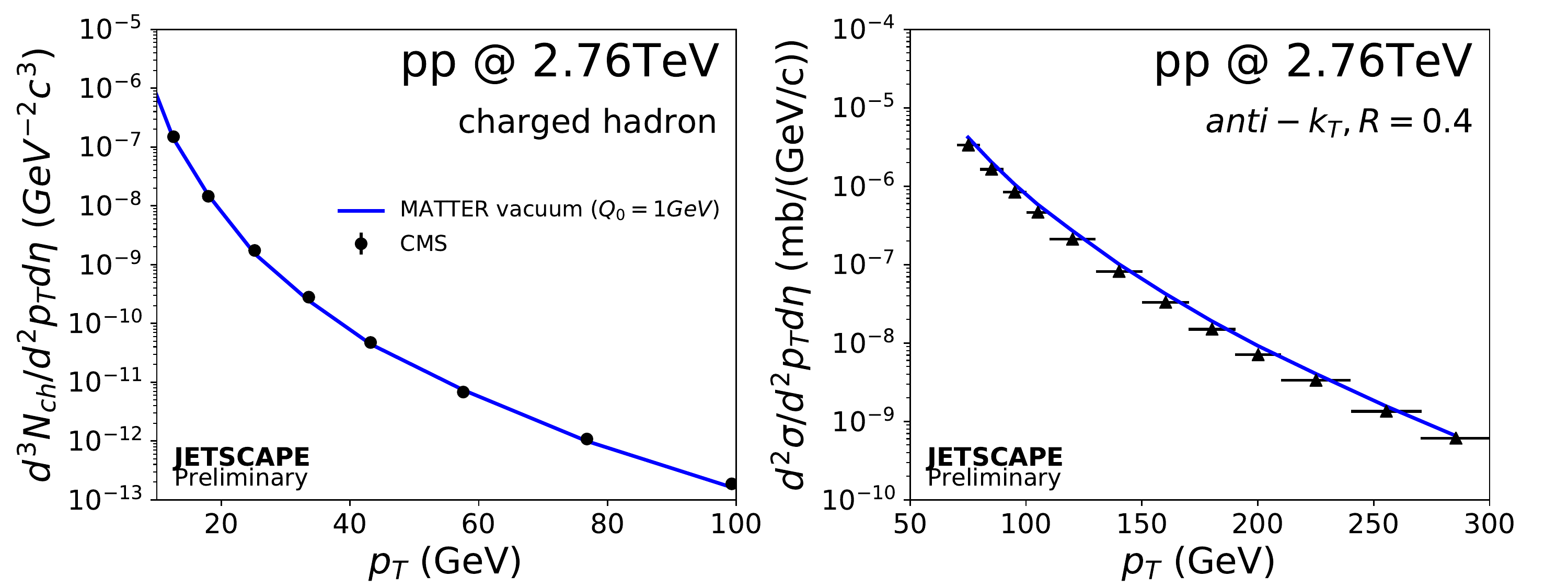}
	\caption[]{Charged hadron differential yield (left) and differential jet cross section (right) at 2.76 TeV pp collisions, calculated using JETSCAPE and compared to measured data~\cite{CMS:2012aa,Khachatryan:2016jfl}.}
	\label{fig:pt_spectra}
\end{figure*}

\begin{figure*}[t]
	\centering
	\includegraphics[width=\textwidth]{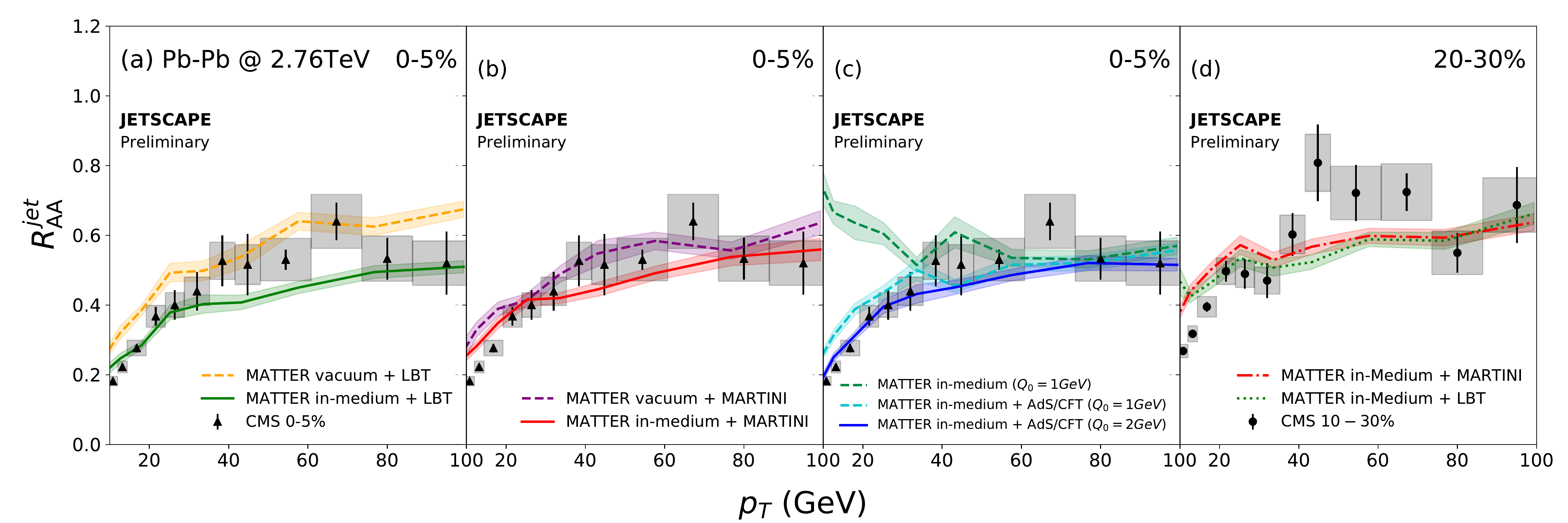}
	\caption[]{Charged hadron $R_{AA}$ for $0$-$5\%$ PbPb collisions at 2.76 TeV measured by CMS~\cite{CMS:2012aa}, compared to JETSCAPE calculations that couple MATTER with (a) MARTINI, (b) LBT, and (c) AdS/CFT. (d): Same calculations for $20$-$30\%$ PbPb collisions.}
	\label{fig:RAA_charged_hadron}
\end{figure*}

Figure \ref{fig:pt_spectra} shows a JETSCAPE calculation for inclusive hadron and jet production in pp collisions at $2.76$ TeV by the MATTER vacuum shower, which is terminated when the virtuality of a given parton is down to $1$ GeV.
The calculation reproduces the measured data well.
Using these comparisons as baseline validation, we then calculated single hadron and jet observables employing three different combinations of energy loss modules, by coupling MATTER  with LBT, MARTINI, or AdS/CFT.
As shown in Figure \ref{fig:RAA_charged_hadron}, the measurements of charged hadron yield suppression, $R_{AA}$, are well-described by all model combinations.

\begin{figure*}[t]
	\centering
	\includegraphics[width=0.9\textwidth]{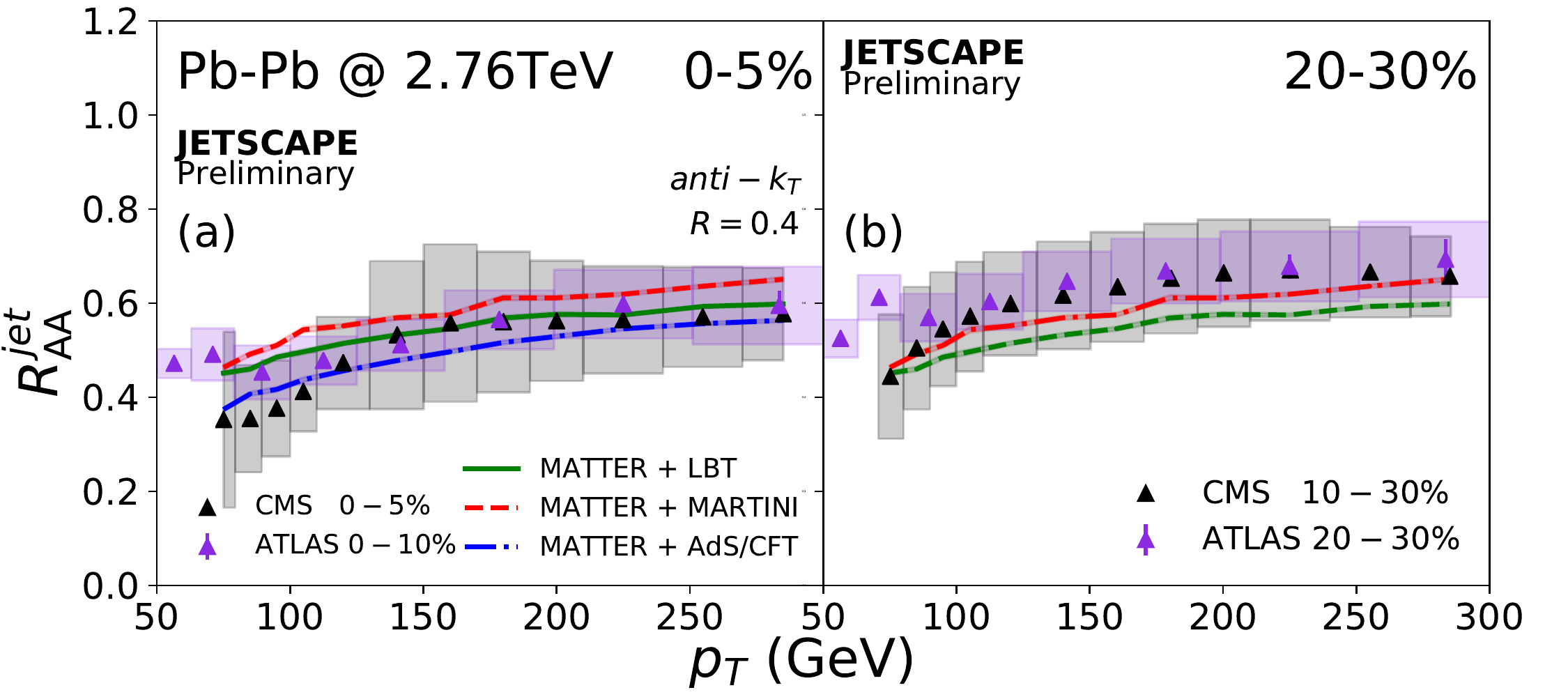}
	\caption[]{Jet $R_{AA}$ calculated using JETSCAPE with three different modules for low virtuality, for (a) $0$-$5\%$ and (b) $20$-$30\%$ PbPb collisions at 2.76 TeV, compared to experimental data~\cite{Khachatryan:2016jfl,Aad:2014bxa}.}
	\label{fig:RAA_jet}
\end{figure*}

Figure \ref{fig:RAA_jet} shows the JETSCAPE calculation of inclusive jet $R_{AA}$, which agrees well with the measured data from the CMS and ATLAS experiments.
The three different combinations of energy loss modules yield compatible results for single hadron and jet observables.

\begin{figure*}[t]
	\centering
	\begin{subfigure}[b]{0.45\textwidth}
		\centering
		\includegraphics[width=\textwidth]{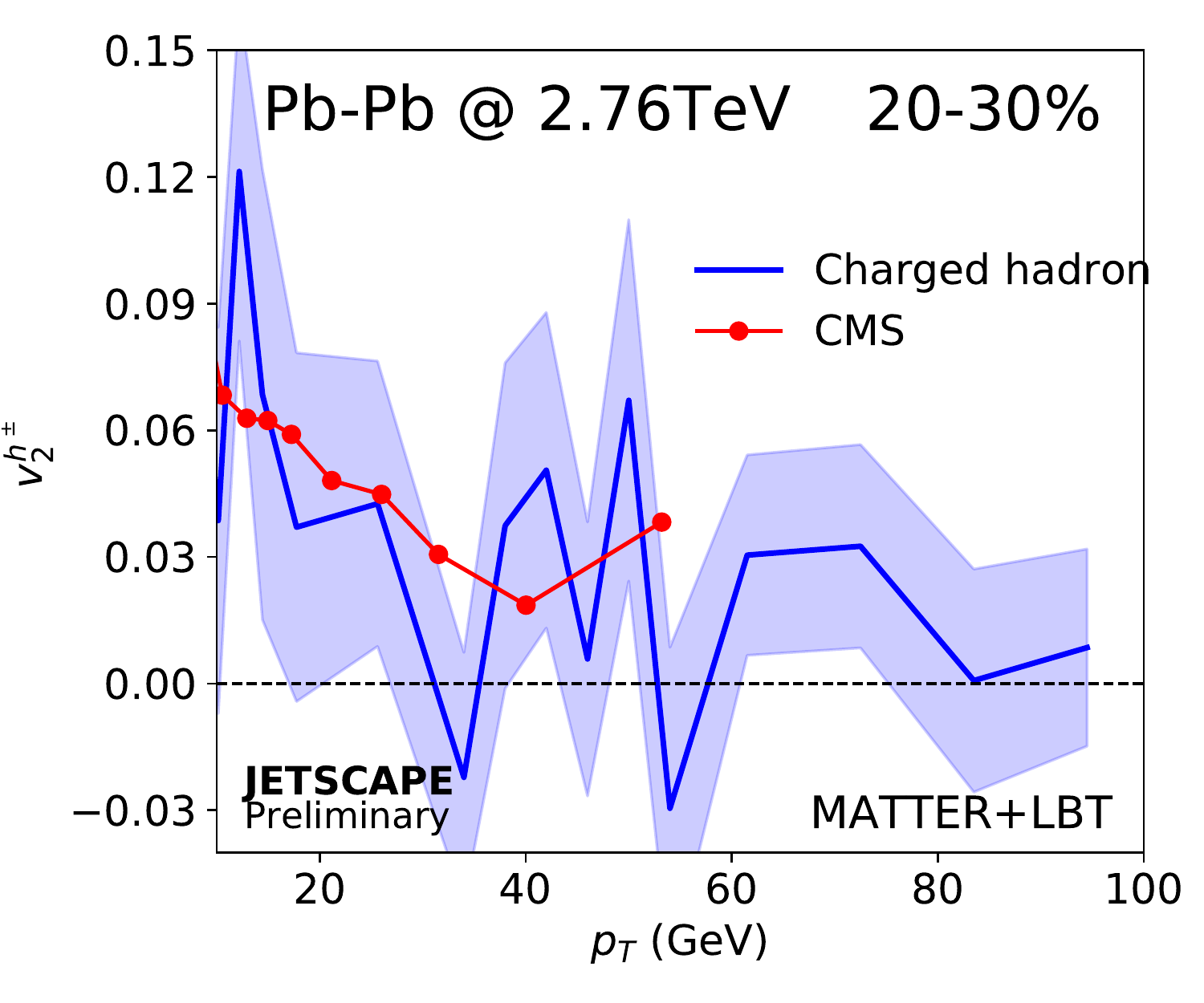}
	\end{subfigure}
	\begin{subfigure}[b]{0.45\textwidth}
		\centering
		\includegraphics[width=\textwidth]{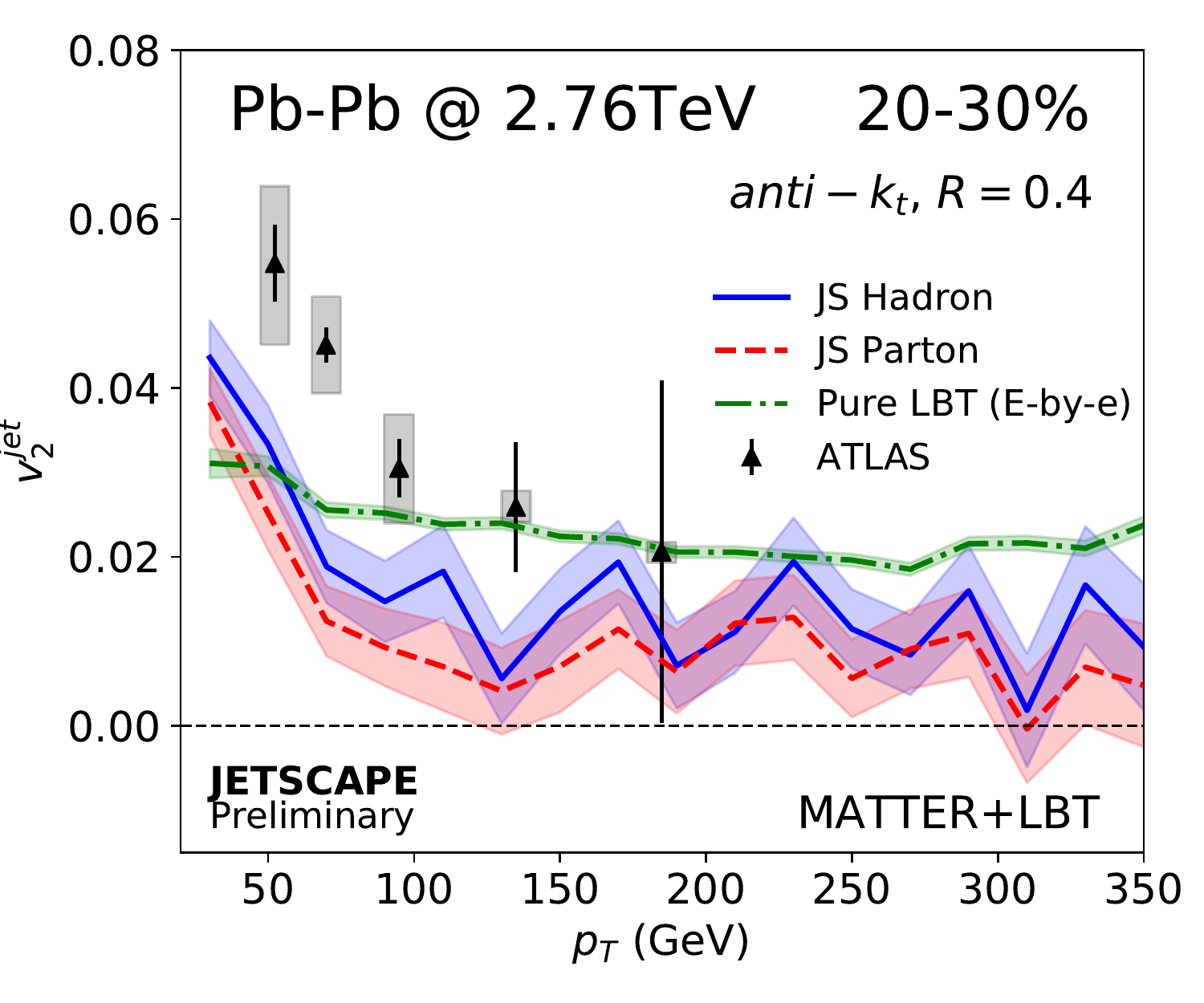}
	\end{subfigure}
	\caption[]{Elliptic flow of (left) charged hadron and (right) jet at $20$-$30\%$ PbPb collisions at 2.76TeV obtained from MATTER with LBT, compared to experimental data~\cite{Chatrchyan:2012ta,Aad:2013sla}.
	}
	\label{fig:v2}
\end{figure*}

Figure \ref{fig:v2} shows the $p_T$ differential elliptic flow coefficients for charged hadron (left) and jets (right) produced in $20$-$30\%$ PbPb collisions at 2.76TeV, representing the correlation between soft and hard sectors.
While the error bands are sizable, especially for charged hadron $v_2$, our results generally capture correctly the $p_T$ dependent strength of the correlation in both cases.
The level of agreement of the pure LBT calculation, which calculates the hydrodynamic background event-by-event, suggests that the agreement of the jet $v_2$ calculation with data would be improved by utilizing a fluctuating medium.

\section{Conclusion}

In this work we have developed a unified approach  to the modeling of jet quenching, combining a set of energy loss models for each distinct stage of the high-energy jet in-medium evolution.
The virtuality-ordered shower for high virtuality partons (MATTER) was coupled to low-virtuality parton evolution models (LBT, MARTINI, and AdS/CFT).
We demonstrated that our approach provides a good description of inclusive hadron and jet suppression, with the different approaches generating consistent results for the interplay between jets and QCD plasma.


\begin{thebibliography}{99}

\bibitem{Kauder:2018cdt} 
  K.~Kauder [JETSCAPE Collaboration],
  arXiv:1807.09615 [hep-ph].
  
\bibitem{Cao:2017zih} 
  S.~Cao {\it et al.} [JETSCAPE Collaboration],
  Phys.\ Rev.\ C {\bf 96}, no. 2, 024909 (2017)
  doi:10.1103/PhysRevC.96.024909
  [arXiv:1705.00050 [nucl-th]].
  
\bibitem{Majumder:2013re} 
  A.~Majumder,
  Phys.\ Rev.\ C {\bf 88}, 014909 (2013)
  doi:10.1103/PhysRevC.88.014909
  [arXiv:1301.5323 [nucl-th]].
  
\bibitem{Cao:2017hhk} 
  S.~Cao, T.~Luo, G.~Y.~Qin and X.~N.~Wang,
  Phys.\ Lett.\ B {\bf 777}, 255 (2018)
  doi:10.1016/j.physletb.2017.12.023
  [arXiv:1703.00822 [nucl-th]].
  
\bibitem{Schenke:2009gb} 
  B.~Schenke, C.~Gale and S.~Jeon,
  Phys.\ Rev.\ C {\bf 80}, 054913 (2009)
  doi:10.1103/PhysRevC.80.054913
  [arXiv:0909.2037 [hep-ph]].

\bibitem{Casalderrey-Solana:2014bpa} 
  J.~Casalderrey-Solana, D.~C.~Gulhan, J.~G.~Milhano, D.~Pablos and K.~Rajagopal,
  JHEP {\bf 1410}, 019 (2014)
  Erratum: [JHEP {\bf 1509}, 175 (2015)]
  doi:10.1007/JHEP09(2015)175, 10.1007/JHEP10(2014)019
  [arXiv:1405.3864 [hep-ph]].
  
\bibitem{Sjostrand:2007gs} 
  T.~Sjostrand, S.~Mrenna and P.~Z.~Skands,
  Comput.\ Phys.\ Commun.\  {\bf 178}, 852 (2008)
  doi:10.1016/j.cpc.2008.01.036
  [arXiv:0710.3820 [hep-ph]].

\bibitem{Shen:2014vra}
  C.~Shen, Z.~Qiu, H.~Song, J.~Bernhard, S.~Bass and U.~Heinz,
  Comput.\ Phys.\ Commun.\  {\bf 199} (2016) 61
  doi:10.1016/j.cpc.2015.08.039
  [arXiv:1409.8164 [nucl-th]].
  
\bibitem{Moreland:2014oya} 
  J.~S.~Moreland, J.~E.~Bernhard and S.~A.~Bass,
  Phys.\ Rev.\ C {\bf 92}, no. 1, 011901 (2015)
  doi:10.1103/PhysRevC.92.011901
  [arXiv:1412.4708 [nucl-th]].
  
\bibitem{CMS:2012aa} 
  S.~Chatrchyan {\it et al.} [CMS Collaboration],
  Eur.\ Phys.\ J.\ C {\bf 72}, 1945 (2012)
  doi:10.1140/epjc/s10052-012-1945-x
  [arXiv:1202.2554 [nucl-ex]].
  
\bibitem{Khachatryan:2016jfl} 
  V.~Khachatryan {\it et al.} [CMS Collaboration],
  Phys.\ Rev.\ C {\bf 96}, no. 1, 015202 (2017)
  doi:10.1103/PhysRevC.96.015202
  [arXiv:1609.05383 [nucl-ex]].
  
\bibitem{Aad:2014bxa} 
  G.~Aad {\it et al.} [ATLAS Collaboration],
  Phys.\ Rev.\ Lett.\  {\bf 114}, no. 7, 072302 (2015)
  doi:10.1103/PhysRevLett.114.072302
  [arXiv:1411.2357 [hep-ex]].
  
\bibitem{Chatrchyan:2012ta} 
  S.~Chatrchyan {\it et al.} [CMS Collaboration],
  Phys.\ Rev.\ C {\bf 87}, no. 1, 014902 (2013)
  doi:10.1103/PhysRevC.87.014902
  [arXiv:1204.1409 [nucl-ex]].
    
\bibitem{Aad:2013sla} 
  G.~Aad {\it et al.} [ATLAS Collaboration],
  Phys.\ Rev.\ Lett.\  {\bf 111}, no. 15, 152301 (2013)
  doi:10.1103/PhysRevLett.111.152301
  [arXiv:1306.6469 [hep-ex]].
  
\end{thebibliography}
\end{document}